\documentstyle[prb,preprint,aps]{revtex}

\begin{document}
\draft
\title{Proportion of frozen local polarization in relaxor ferroelectrics}
\author{Zhi-Rong Liu
\footnote{Electronic address: zrliu@phys.tsinghua.edu.cn}}
\address{Department of Physics, Tsinghua University, 
Beijing 100084, People's Republic of China}
\author{Yong Zhang}
\address{Beijing Fine Ceramics Laboratory, 
Institute of Nuclear Energy Technology, Tsinghua University, 
Beijing 100084, People's Republic of China}
\author{Bing-Lin Gu} 
\address{Center for Advanced Study, Tsinghua University, 
Beijing 100084, People's Republic of China}
\author{Xiao-Wen Zhang}
\address{State Key Laboratory of New Ceramics and Fine Processing, 
Department of Materials Science and Engineering, 
Tsinghua University, Beijing 100084, People's Republic of China}
\maketitle

\begin{abstract}

A Landau-type phenomenological cluster theory was presented to 
model the freezing process of local polarization in relaxors. 
Based on the theory, 
the proportion of frozen polarization in 
Pb(Mg$_{1/3}$Nb$_{2/3}$)O$_3$-PbTiO$_3$ (PMNT) was calculated from the 
experiment of dielectric nonlinearity. 
The local polarization was shown to freeze continuously in a cooling 
process. The amount of frozen polarization increases with increasing 
the measuring frequency.

\end{abstract}

\pacs{PACS:  77.22.Ej, 77.80.-e, 77.22.Ch, 77.84.Lf }

%64.60.Cn, 82.20.Mj, 64.60.My, 82.20.Wt 77.22.G, 77.80, 77.84.L
% 77.22.Ej Polarization and depolarization
%77.80.-e Ferroelectricity and antiferroelectricity
%77.22.Ch Permittivity (dielectric function)
%77.84.Lf Composite materials
%77.84.-s Dielectric, piezoelectric, ferroelectric, and
%         antiferroelectric materials (for nonlinear
%         optical materials, see 42.70.M)

\vspace{2mm}

%\end{titlepage}  

%\section{Introduction}

Relaxor ferroelectrics (relaxors) have been showing great importance both 
for the fundamental solid state science and the applications to 
advanced technology.\cite{1,2,3} Relaxors experience no macroscopic 
phase transition at zero electric field until very low temperatures. 
However,  local polarization exists at much higher temperatures, which is 
widely believed to be responsible for the special characteristics of relaxors. 
The concept of ``polar micro region" in relaxors was firstly 
presented by Smolenskii in the chemical inhomogeneity theory.\cite{4} 
The presence 
of polar microregions was later confirmed by the experiments 
on the non-linear variation of the optical refractive index, 
the thermal strain, and the thermal expansion coefficient.\cite{1,5,6,7} 
The deviation from the paraelectric Curie-Weiss 
behavior of the dielectric permittivity also suggested the presence 
of polar regions. Cross presented in a superparaelectric model that 
the polar microregions (represented by independent dipoles) are able 
to switch between the equivalent orientation states and gave 
an appropriate picture of the nature of the micro polarization at 
high temperatures.\cite{1} When the temperature decreases, the coupling 
between the polar microregions controls the kinetics of the fluctuations 
and the system is frozen into a polar-glassy state.\cite{8,9} 
Gui {\it et al.} used the Monte Carlo method to simulate the freezing process 
in relaxors and showed that some dipoles are slowed or 
frozen due to the interactions.\cite{10} A special phase transition of 
ergodic space shrinking in succession was also proposed according to 
the freezing process.\cite{11} Kleemann {\it et al.}, however, proposed that 
the polarization is frozen due to the quenched random electric fields.
\cite{12,13}

Some works have been done to quantitatively analyze the freezing process 
of local polarization. 
Qian and Bursill developed a phenomenological 
theory to describe the interaction of polar domains and to simulate 
the dielectric relaxation and phase transition of relaxors.\cite{14} Nambu 
and Sugimoto proposed a Landau type mean field theory by considering a 
gradual condensation of local polarization and confirm a picture 
of diffuse phase transition in relaxors.\cite{15} In this paper, we 
propose a Landau-type phenomenological cluster theory of relaxors 
and calculate the proportion of frozen polarization from the experiments 
of dielectric nonlinearity.

%\section{Methods}

A phenomenological free energy is defined as 
\begin{equation}
F=2^{-1}N\alpha(T) P^2+4^{-1}N\alpha_{11}P^4
  +2^{-1}\gamma P^2\sum\limits_{i}n_iP_i^2
  +2^{-1}\sum\limits_{i}n_i\alpha_i(T) P_i^2
  +4^{-1}\alpha'_{11}\sum\limits_{i}n_i P_i^4 ,
\end{equation}
where $P$ is the uniform polarization and $P_i$ is the $i$-th frozen local 
polarization. $N$ is the total number of lattice sites, and $n_i$ is 
the lattice-site number of the $i$-th local polarization. $\alpha$ 
and $\alpha_i$ are written as 
\begin{eqnarray}
\alpha(T)=(T-T_0)/\varepsilon_0C, \\
\alpha_i(T)=(T-T_i)/\varepsilon_0C_i,
\end{eqnarray}
and $\alpha_{11}$ and $\alpha'_{11}$ are constants as the conventional 
Landau theory. The couplings between the global polarization and the local ones 
are written as $\gamma P^2 P_i^2$ 
from the symmetry consideration. 

The equilibrium values of the frozen polarization $\{P_i\}$ are determined by 
minimizing the free energy $F$ in Eq. (1). Minimizing $F$ with 
respect to $\{P_i\}$ gives
\begin{equation}
\partial F/\partial P_i=0,
\end{equation}
i.e., 
\begin{equation}
\gamma P^2 n_iP_i
  +n_i\alpha_i(T) P_i
  +\alpha'_{11}n_i P_i^3=0 .
\end{equation}
Thus $P_i$ is solved from Eq. (5) as
\begin{equation}
P_i^2=-\alpha_i(T)/\alpha'_{11}-\gamma P^2/\alpha'_{11}.
\end{equation}
Define $P_{i0}$ as
\begin{equation}
P_{i0}^2=-\alpha_i(T)/\alpha'_{11}=(T_i-T)/\varepsilon_0C_i\alpha'_{11},
\end{equation}
and then $P_i$ can be rewritten as
\begin{equation}
P_i^2=P_{i0}^2-\gamma P^2/\alpha'_{11}.
\end{equation}
When relaxors stays in a unpolar state ($P$=0), the frozen polarization 
$P_i$ is equal to $P_{i0}$.
Substituting Eqs (7) and (8) into Eq. (1) yields the free energy
\begin{equation}
\overline {F}=2^{-1}NP^2
         \left[ \alpha(T)+\gamma\frac{1}{N}\sum\limits_{i}n_i P_{i0}^2 \right]
  +4^{-1}NP^4
         \left[\alpha_{11}-\frac{\gamma^2}{\alpha'_{11}}\cdot 
                         \frac{1}{N}\sum\limits_{i}n_i \right]
   +{\rm const}.
\end{equation}
Introduce 
\begin{equation}
q(T)=\frac{1}{N}\sum\limits_{i}n_i P_{i0}^2
\end{equation}
and
\begin{equation}
n(T)=\frac{1}{N}\sum\limits_{i}n_i,
\end{equation}
and then Eq. (9) can be expressed as
\begin{equation}
\overline {F}=2^{-1}Na(T)P^2+4^{-1}Nb(T)P^4+{\rm const},
\end{equation}
where
\begin{eqnarray}
a(T)=\alpha(T)+\gamma q(T), \\
b(T)=\alpha_{11}-\frac{\gamma^2}{\alpha'_{11}}n(T).
\end{eqnarray}
For relaxors, the sum of local polarizations is equal to zero, while the 
sum of the square of local polarizations is not zero that can be measured in 
the nonlinear behaviors of optical refractive index and the thermal strain, 
etc.\cite{1,5,6,7} So $q(T)$ is the order parameter to represent the appearence of local 
polarizations. $n(T)$ is the proportion 
of the lattice site of frozen polarization.
It should be noted that Eq. (7) is valid only for $T<T_i$ (for $T>T_i$, 
$P_{i0}$=0, 
i.e., the $i$-th local polarization is not frozen), so the summarization in Eqs. 
(9-11) can be conducted only 
on the nonzero local polarization ($P_{i0}^2>0$), and $n(T)$ and $q(T)$ 
vary with temperatures.

The dielectric susceptibility is easily calculated from the thermodynamic 
relation between the dielectric field, 
\begin{equation}
E=\frac{1}{N}\cdot\frac{\partial \overline {F}}{\partial P}
  =a(T)P+b(T)P^3,
\end{equation}
and the uniform polarization $P$, i.e., 
\begin{equation}
\frac{1}{\varepsilon(T)}=\left.\frac{\varepsilon_0 E}{P}\right|_{E=0}
=\varepsilon_0a(T)=\frac{T-T_0}{C}+\varepsilon_0\gamma q(T).
\end{equation}
It can be seen that the dielectric susceptibility deviates from the 
Curie-Weiss behavior due to the existence of the local order parameter 
$q(T)$. 
Schmitt and Kirsch gave the similar relation of Eq. (16) by simple 
phenology in (Pb,La)(Zr,Ti)O$_3$ (PLZT) system.\cite{16} Nambu and 
Sugimoto derived the same relation on the basis of a more general 
mean field theory and explained the difference of the phase transitions 
in PMN and Pb(Sc$_{1/2}$Ta$_{1/2}$)O$_3$ (PST).\cite{15} 
The purpose of this works was to investigate $n(T)$, but not $q(T)$. 
So we consider the dielectric nonlinearity, i.e., the susceptibility 
corresponding to a nonzero electric field $E$, 
\begin{equation}
\varepsilon_E(T)=\frac{P}{\varepsilon_0 E}
              =\frac{1}{\varepsilon_0}\cdot\frac{1}{a(T)+b(T)P^2}
 \approx \frac{1}{\varepsilon_0a(T)}-\frac{b(T)}{\varepsilon_0a^2(T)}P^2
=\varepsilon(T)-\varepsilon_0^3E^2\cdot\varepsilon^2(T)\varepsilon_E^2(T)
\cdot b(T) .
\end{equation}
According to Eqs (17) and (14), $n(T)$ can be obtained as
\begin{equation}
n(T) = \frac{\alpha'_{11}}{\gamma^2}\left[
\alpha_{11}+\frac{1}{\varepsilon_0^3E^2}
\frac{\varepsilon_E(T)-\varepsilon(T)}{\varepsilon^2(T)\varepsilon_E^2(T)}
\right].
\end{equation}
The value of $\alpha_{11}$ can be obtained from the experiment data 
at high temperatures by setting $n(T)=0$ in Eq. (18) 
since there is no frozen polarization 
at high temperatures. So, by using Eq. (18), the proportion of frozen polarization, $n(T)$, 
can be determined except a constant coefficient $\alpha'_{11}/\gamma^2$ 
from the experiment of dielectric nonlinearity. 

We take a typical and well-known relaxors, 
Pb(Mg$_{1/3}$Nb$_{2/3}$)O$_3$-PbTiO$_3$ 
(PMNT), as an example to analyze the proportion of frozen polarization. 
The samples were prepared by two-stage calcination method. 
The magnesium niobate is synthesized by calcining 
MgCO$_3\cdot$Mg(OH)$_2\cdot$6H$_2$O and Nb$_2$O$_5$ at 1000$^\circ C$ for 6 h. 
The columbite phase was then mixed with lead oxide and titanium oxide to 
form the composition 0.96PMN-0.04PT. Excess PbO (0.3 wt\%) was added to 
compensate for PbO loss during heat treatment. The mixture was 
ball-milled and calcined again at 800$^\circ C$ for 2 h. A 10 mm uniaxial 
steel die was employed to produce green pellets from the calcined powder, 
using a pressure of 100 MPa. The pressed pellets were subsequently 
sintered in a covered alumina crucible at 1200$^\circ C$ for 1 h. 
After sintering, the pellets were polished to a thickness of 0.6 mm and 
silver paste was applied and fired at 600$^\circ C$ to achieve a 
conductive and adherent coating. The dielectric susceptibility was 
measured at various frequencies with an LCR precision meter (Model 
HP 4284A, Hewlett Packard, Palo Alto, CA) remotely controlled through a 
desktop computer. A temperature chamber (Model 2300, Delta Design, 
San Diego, CA) was interfaced to the computer to allow measurement of 
dielectric properties at various temperatures. The amplitude of the ac 
measurement field was 0.05 and 0.25 kV/cm.

%\section{Results and Discussions}

An illustration of the temperature dependence of the dielectric 
response at a low frequency 100 Hz is shown in Fig. 1(a), and the 
corresponding calculated proportion of frozen polarization is given in 
Fig. 1(b). The curves in Fig. 1(a) demonstrate typical dielectric nonlinear 
behavior with the magnitude of the susceptibility increasing with 
increasing the field amplitude and the maximum shifting to lower 
temperatures, which is consistent with the previous observations.\cite{17} The 
curve of the proportion of frozen polarization, $n(T)$, is determined with 
an arbitrary factor. It can be seen that $n(T)$ is equal to zero at 
high temperatures, which means that the local polarization is all 
dynamic.\cite{1,14} When the temperature decreases to approach the 
temperature of the susceptibility maximum ($T_m$), 
$n(T)$ starts to increase, i.e., some local polarization is frozen. 
After crossing $T_m$, $n(T)$ increases 
rapidly with decreasing temperature, and an extrapolation of the 
slope at the inflection point to zero yielded a critical temperature 
$T_p=8.3$$^\circ C$. 
The monotonous increasing of $n(T)$ with decreasing temperature in Fig. 1(b) shows 
that the calculation is physically realistic. 

In the works of other researchers,\cite{9,16} the calculated 
quantity is the local order parameter $q(T)$ that is determined from 
the temperature dependence of zero-field susceptibility [see Eq. (16)]. 
The increase of $q(T)$ with decreasing temperature is ensured by the 
decrease of susceptibility at low temperatures.  However, the quantity 
investigated here, $n(T)$, is calculated from the nonlinear effect 
[see Eq. (18)]. 
It should be noted that the dielectric nonlinearity is the strongest when 
the temperature is near $T_m$ while the proportion of frozen polarization 
is not large at that temperature range. It seems to be conflict 
with the thought that the frozen polarization is the origin 
of the dielectric nonlinearity. The key to answer this problem is that the 
nonlinearity is approximately proportional to the forth power of the susceptibility 
[see Eq. (17)]. The susceptibility reaches the maximum at $T_m$, so the 
nonlinear is strong near $T_m$. This can also explain the weakening of 
the nonlinearity when the proportion of frozen polarization 
increases in further decreasing the temperature. 

Another feature of Fig. 1 is that the frozen polarization, i.e., the 
local polarization which experiences a phase transition and loses 
the ergodicity, does not 
appear suddenly, but increases continuously, which is accompanied with 
the decreasing of the susceptibility. It is a kind of phase transition 
where the ergodic space shrinks in succession.\cite{11}

A few sentences can be presented here to explore the origin 
of the effect of dielectric amplitude on the susceptibility. From Eq. (8) 
one knows that the magnitude of frozen polarization decreases 
with increasing external field. In other words, the driving forces on 
the polarization is enhanced when the ac field amplitude increases, 
so the frozen polarization is forced to flip faster, and 
some frozen polarization 
is unfrozen and gives contribution 
to the polarization process. Thus the susceptibility increases with 
increasing the field amplitude. 

Figure 2 shows the curves of frozen polarization proportion at 
different measuring frequencies. When the frequency increases, the curve 
of the proportion shifts slightly towards higher temperatures. It implies 
that the time scale of polarization flipping is shortened when the frequency 
increases, so more polarization cannot reach the equilibrium states in 
the observation time, i.e., more polarization is frozen.
The slight increase of frozen polarization 
would result in the decreasing of susceptibility at higher frequency, 
which is known as the frequency dispersion in relaxors. The 
curve of $T_p$ [$T_p$ is defined in Fig. 1(b)] is depicted in Fig. 3 
together with the curve of $T_m$ for comparison. It clearly demonstrates 
the increasment of $T_p$ with increasing frequency. 
And we can see that $T_m$ increases more rapidly than $T_p$ does.

There are several points that should be discussed here. 
Firstly, the couplings between local polarization are not included 
in Eq. (1). However, if the couplings between polarization are 
considered, through the similar solving procedure as that in Ref. 15, 
the results presented in this paper is still valid by redefining the 
parameters. Secondly, we only consider the forth-power terms in the 
free energy and ignore the influence of the higher terms in this works. 
And the coefficients $\alpha_{11}$, $\alpha_{11}'$ and $\gamma$ are 
considered as temperature-independent. All these defects restrict the 
applications of the theory. At last, in this paper the local properties 
are included within the Landau expansion by introducing the local changing 
quantities $n_i$. Consequently, one expects the appearance of a distribution 
for $n_i$ which is missed in the model. As the result all quantities should be 
averaged on such a distribution.

%\section{Summaries}

In summary, a Landau-type phenomenological cluster theory 
of relaxors is proposed in this paper to evaluate the influence 
of frozen polarization. The proportion of frozen polarization in PMNT 
is calculated from the experimental data of dielectric nonlinearity. 
It is shown that the frozen polarization starts to appear when the 
temperature decreases near to $T_m$, and increases rapidly after 
crossing $T_m$. When the ac field frequency increases, the 
amount of frozen polarization arises too.

%\section*{Acknowledgment}

This work was supported by the Chinese National Science Foundation 
(Grant NO. 59995520) and State Key Program of Basic Research 
Development (Grant No. G2000067108).

%Figure.1
\begin{figure}[tbp]
\caption{(a) The dielectric permittivity of PMNT as functions 
of temperature at various drive amplitudes when the frequency is fixed 
as 100 Hz. (b) The proportion of frozen polarization (arbitrary unit) 
in PMNT calculated from the data in Fig. 1(a) by Eq.(18). }
\end{figure}

%Figure.2
\begin{figure}[tbp]
\caption{The proportion of frozen polarization in PMNT for different 
measurement frequencies.  }
\end{figure}

%Figure.3
\begin{figure}[tbp]
\caption{The curves of $T_m$ (temperature of susceptibility maximum) 
and $T_p$ (defined in Fig. 1) as functions of measurement frequency. }
\end{figure}


\begin{references}
\bibitem{1} L. E. Cross, Ferroelectrics {\bf 76}, 241 (1987).

\bibitem{2} Z. G. Ye, Key Eng. Mater. {\bf 155-156}, 81 (1998).

\bibitem{3} G. H. Haertling, J. Am. Ceram. Soc. {\bf 82}, 797 (1999).

\bibitem{4} G. A. Smolenskii, J. Phys. Soc. Japan {\bf 28 suppl.}, 26 (1970).

\bibitem{5} G. Burns and F. H. Dacol, Solid State Commun. {\bf 48}, 
853 (1983).

\bibitem{6} P. Bonneau, P. Garnier, G. Calvarin, E. Husson, J. R. Gavarri, 
A. W. Hewat, and A. Morell, J. Solid State Chem. {\bf 91}, 350 (1991).

\bibitem{7} H. Arndt and G. Schmidt, Ferroelectrics {\bf 79}, 149 (1988).

\bibitem{8}  D. Viehland, S. J. Jang, L. E. Cross, and M. Wuttig, 
J. Appl. Phys. {\bf 68}, 2916(1990).

\bibitem{9} D.  Viehland, S. J. Jang, L. E. Cross and M. Wuttig, 
Phys. Rev. B {\bf 46}, 8003 (1992).

\bibitem{10} H. Gui, B. L. Gu, and X. W. Zhang, Phys. Rev. B {\bf 52}, 
    3135 (1995).

\bibitem{11} X. W. Zhang, H. Gui, Z. R. Liu, and B. L. Gu, Phys. Lett. A 
{\bf 251}, 219 (1999).

\bibitem{12} V. Westphal, W. Kleemann, and M. D. Glinchuk, Phys. Rev. 
Lett. {\bf 68}, 847 (1992).

\bibitem{13} Kleemann, Int. J. Mod. Phys. B {\bf 7}, 2469 (1993).

\bibitem{14} H. Qian and L. A. Bursill, Int. J. Mod. Phys. B {\bf 10}, 
2007 (1996).

\bibitem{15} S. Nambu and K. Sugimoto, Ferroelectrics {\bf 198}, 11 (1997).

\bibitem{16} H. Schmitt and B. Kirsch, Ferroelectrics {\bf 124}, 
225 (1991).

\bibitem{17} A. E. Glazounov, A. K. Tagantsev, and A. J. Bell, Phys. Rev. B 
{\bf 53}, 11281 (1996).

\end{references}
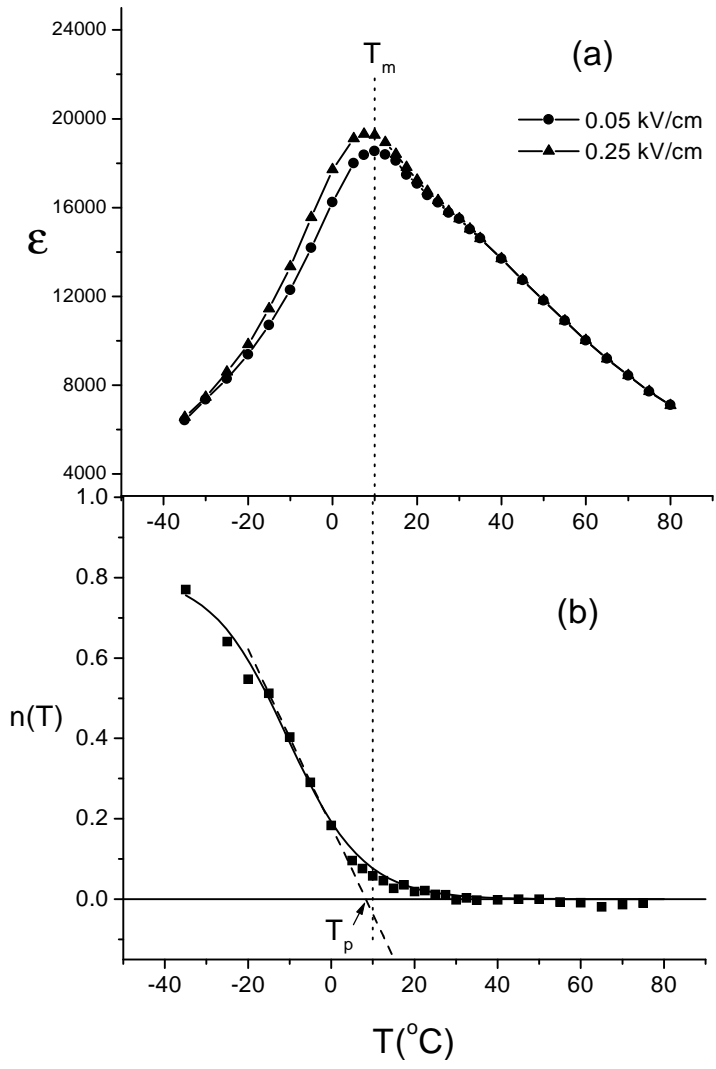
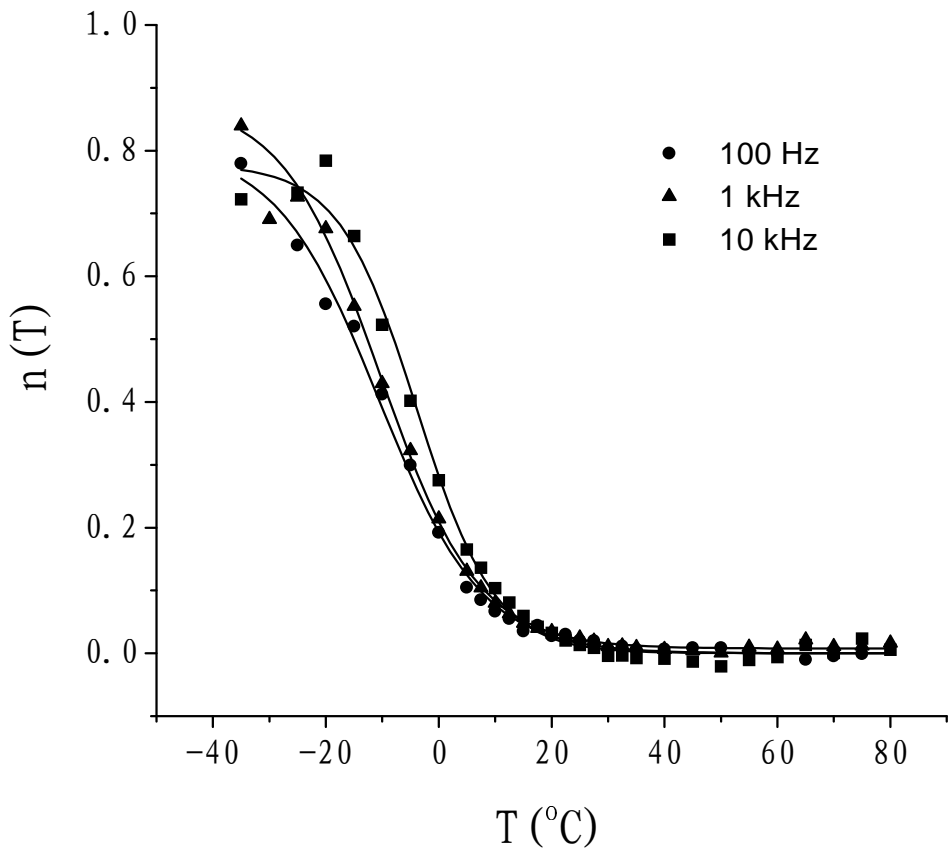
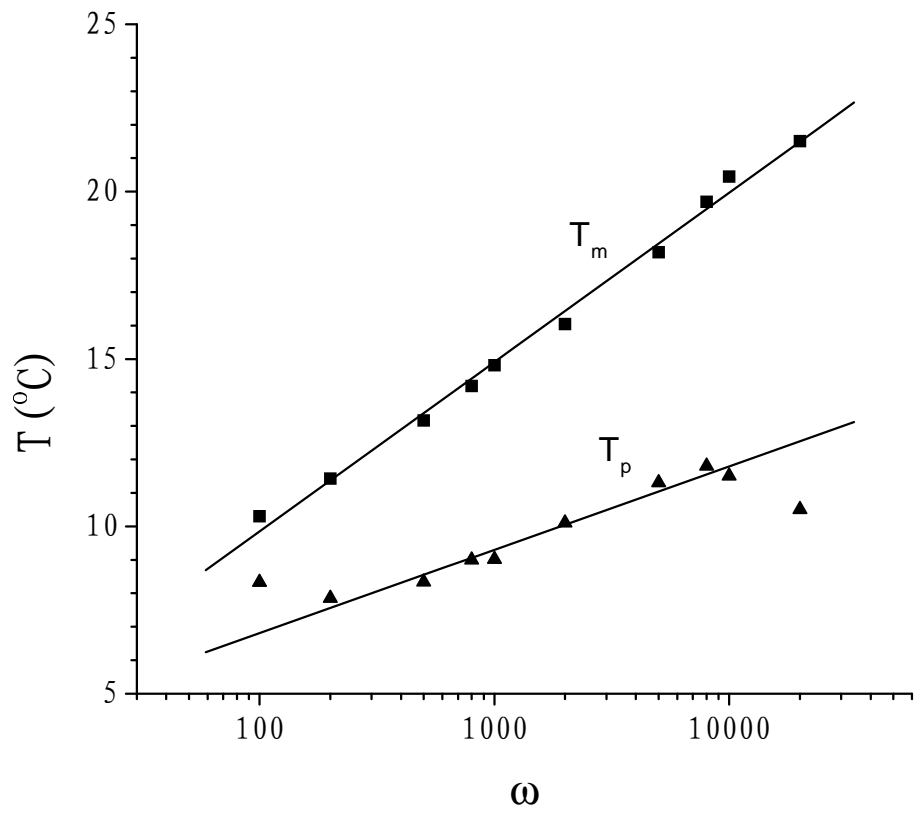
\end{document}